\documentclass[flushrt,10pt,preprint]{aastex}
\usepackage{txfonts}

\newcommand\STAR{HD~200775}
\newcommand\ngc{NGC~7023}
\newcommand\PN{NGC~7027}
\newcommand\ON{Orion Nebula}
\newcommand\OB{Orion Bar}
\newcommand\tCO{$\theta^1$C~Orionis}

\newcommand\HI{\ion{H}{1}}
\newcommand\HII{\ion{H}{2}}
\newcommand\hh{{\rm H}$_2$}
\newcommand\HCO{{\rm HCO}$^+$}
\newcommand\ppix{\ensuremath{\mbox{pixel}^{-1}}}
\newcommand\ttc{{\rm [}2.18~{\rm \micron]}$-${\rm [}3.29~{\rm \micron]}}
\newcommand\Teff{\ensuremath{T\!_{\rm eff}}}

\begin{document}
\slugcomment{\sl to appear in \it the Astrophysical Journal}
\title{\textsc{\textbf{\boldmath
Spatial Separation of the 3.29~{\bf\micron} Emission Feature\\
and Associated 2~{\bf\micron} Continuum in \ngc}}}
\author
{Jin H. An\altaffilmark{1} and K. Sellgren\altaffilmark{2}}
\affil
{Department of Astronomy, Ohio State University,
140 West 18th Avenue, Columbus, OH 43210;}
\email
{jin@ast.cam.ac.uk, sellgren@astronomy.ohio-state.edu}
\altaffiltext{1}
{Current Address: Institute of Astronomy, University of Cambridge,
Madingley Road, Cambridge, CB3 0HA, UK}
\altaffiltext{2}
{Visiting Astronomer at the Infrared Telescope Facility, which is
operated by the University of Hawaii under a cooperative agreement
with the National Aeronautics and Space Administration (NASA).}

\begin{abstract}
We present a new 0\farcs9 resolution 3.29~\micron\ narrowband image of
the reflection nebula \ngc. We find that the 3.29~\micron\ infrared
emission feature (IEF) in \ngc\ is brightest in narrow filaments
northwest of the illuminating star \STAR. These filaments have been
previously seen in images of extended red emission, $K'$,
near-infrared \hh\ emission lines, the 6.2~\micron\ and 11.3~\micron\
IEFs, and \HCO. We also detect 3.29~\micron\ IEF emission faintly but
distinctly between the filaments and the star. The 3.29~\micron\ IEF
image is in marked contrast to narrowband continuum images at
2.09~\micron, 2.14~\micron, and 2.18~\micron, which show an extended
emission peak midway between the filaments and the star, and much
fainter emission near the filaments. The \ttc\ color shows a wide
variation, ranging from colors of 3.4-3.6~mag at the 2~\micron\
continuum emission peak to a color of 5.5~mag in the 3.29~\micron\ IEF
emission filaments. A color of 5.5~mag is 1~mag higher than any
published value in reflection nebulae. We observe \ttc\ to increase
smoothly with increasing projected distance from the illuminating star
in \ngc, up until the brightest 3.29~\micron\ IEF filament, suggesting
that the main difference between the spatial distributions of the
2~\micron\ continuum and the 3.29~\micron\ IEF emission is related to
the incident stellar flux. The 3.29~\micron\ IEF is widely attributed
to an aromatic C\sbond H stretch, while the 2~\micron\ continuum
emission may be due to transient heating of tiny grains, fluorescence
of aromatic molecules, or photoluminescence of larger grains. Our
results suggest that the 3.29~\micron\ IEF carriers are likely to be
distinct from, but related to, the 2~\micron\ continuum emitters. Our
finding also imply that, in \ngc, the 2~\micron\ continuum emitters
are mainly associated with \HI, while the 3.29~\micron\ IEF carriers
are primarily found in warm \hh, but that both the 2~\micron\
continuum emitters and the 3.29~\micron\ IEF carriers can survive in
\HI\ or \hh.
\end{abstract}


\section{Introduction\label{intro}}

The 3.29~\micron\ infrared emission feature (IEF), discovered in the
planetary nebula \objectname[]{NGC~7027} \citep*{MSR75}, is the
shortest wavelength IEF among what used to be known as ``the
unidentified infrared features.'' The IEFs are observed at
3.29~\micron, 6.2~\micron, 7.7~\micron, 8.6~\micron, 11.3~\micron, and
12.7~\micron. \citet{DW81} were the first to suggest that the
3.29~\micron\ IEF is due to the stretching mode of an aromatic
($sp^2$) C\sbond H. This identification is now widely accepted
\citep*[see reviews by][]{ATB89,PL89,Se90,Se94,PCG96,To97,Ge97,Sa99}.

The 3.29~\micron\ IEF emission in visual reflection nebulae, such as
\objectname[]{NGC~7023}, has always been observed to be accompanied by
near-infrared (NIR) continuum emission at 2-4~\micron\
\citep*{SWD83,Se84,JDG90a,JDG90b,SWA96,Joblin96}, which is not due to
reflected starlight \citep*{SWD92}. The temperature of dust grains in
equilibrium with the incident stellar radiation for these visual
reflection nebulae is observed to be far too low to produce any
detectable NIR radiation
\citep*{HTG80,WGH81,CSW87,SLW90,Ca91,YO02}. Observations of \ngc\ at
55-400~\micron, in particular, are best fit by an equilibrium dust
temperature of $\sim$50~K \citep{WGH81,Ca91}, in clear disagreement
with the characteristic temperature, $\sim$1000~K, of the NIR
continuum emission \citep{SWD83,Se84}. This NIR continuum emission in
visual reflection nebulae has been attributed to single stellar
photons that transiently heat tiny (1~nm radius) grains to high
temperature \citep{SWD83,Se84}; to vibrational or electronic
fluorescence from polycyclic aromatic hydrocarbon (PAH) molecules
\citep*{LP84,ATB85,ATB89}; or to photoluminescence of larger
hydrogenated amorphous carbon (HAC) grains or carbon nanoparticles
\citep*{DW88,Du88,Du01}.

When an ultraviolet (UV) source, such as a hot star, illuminates a
molecular cloud, it forms a photodissociation region (PDR; also
referred to as a photon-dominated region) where absorption of UV
photons in the Lyman-Werner bands (91.2-110~nm) of molecular hydrogen
(\hh) dissociate \hh. These same photons also pump fluorescent
quadrupole ro-vibrational \hh\ emission lines, and thus fluorescent
\hh\ emission delineates the \HI\,/\hh\ transition region in a PDR.
Observations of the \objectname[]{Orion Bar}, the PDR adjoining the
southeastern ionization front of the \objectname[]{Orion Nebula}
\citep{STN90,TMW93}, and of the planetary nebula \PN\ \citep{GSH93},
have shown that the 3.29~\micron\ IEF emission is strongest between
the \HII\ region ionization front (traced by H recombination lines)
and the \HI\,/\hh\ dissociation front (traced by fluorescent \hh\
emission lines).

Here we present a new high-resolution (0\farcs9) narrowband
3.29~\micron\ image of the bright reflection nebula \ngc, and compare
the spatial distribution of the 3.29~\micron\ IEF emission to those of
other prominent NIR emission components such as the NIR continuum
emission near 2~\micron\ and the 1--0 S(1) \hh\ emission line at
2.12~\micron\ \citep[this paper;][]{LFG96}.

\ngc, at a distance of $430^{+160}_{-90}\ \mbox{pc}$
\citep{hipparcos}, is one of the best-studied reflection nebulae and
is also one of the first objects in which the NIR continuum emission
accompanying the 3.29~\micron\ IEF emission was detected
\citep{SWD83}. It is a part of a much larger molecular cloud,
illuminated by the pre-main sequence Herbig Be star
\objectname[]{HD~200775}, with an effective temperature \Teff\ =
17,000~K \citep*{SSY72,BBK82}. The $K'$ (broadband 2.1~\micron) image
of \ngc\ \citep[Fig.~\ref{f1}; see also][]{SWD92}, which contains a
mixture of 2~\micron\ continuum emission, 2~\micron\ scattered
starlight, and \hh\ line emission, is dominated by extended nebulosity
northwest of \STAR, and by two filamentary structures, northwest and
south/southwest of the star. Various previous imaging studies show
that these two groups of filaments are also seen in the extended red
emission \citep*{WS88,WGS91}, fluorescent \hh\ emission lines
\citep*{LFG96,TUS00}, IEFs at 6.2~\micron\ and 11.3~\micron\
\citep{CLA96}, and \HCO\ emission \citep*{FMN96}. These filaments are
believed to be the interface between the molecular cloud, traced by CO
emission \citep{GPK98}, and photodissociated \hh, traced by the 21~cm
\HI\ line, surrounding the star \citep{FMN96}. The filaments may
represent high density (as high as $n\sim 10^6\ \mbox{cm$^{-3}$}$)
clumps embedded in somewhat lower density gas ($n\sim 10^4\
\mbox{cm$^{-3}$}$) at the \HI\,/\hh\ interface
\citep*{FMN96,LFG96,MSH97,MSD99,TUS00}.

\section{Observations and Data Reduction\label{obs}}

Table~\ref{obs_log} summarizes the observations for the data presented
in this paper.

We obtained an image of \ngc\ at 3.29~\micron\ with the 3~m NASA
Infrared Telescope Facility (IRTF) at Mauna Kea, Hawaii, during an
IRTF service observing run. The total integration time on source is 12
minutes. The central position of the field is 26\arcsec~W 32\arcsec~N
of the illuminating star \STAR, and was selected so as to include all
the northwest filaments and the central star. The data were reduced in
the usual way: sky subtraction and flat field correction. The sky was
observed in an ABBA pattern, chopping between the target field and a
field 5\arcmin~N every 20 seconds. Sky images were used for flat
fields as well. The final image was formed by combining 40 individual
frames using a median filter to increase the signal-to-noise ratio
(S/N), and an average $\sigma$-clipping algorithm was used during the
combination to reject bad pixels. The final image shows a strong ghost
image (instrumental reflected light) of \STAR\ at the diagonally
opposite corner (near 48\arcsec~W 60\arcsec~N) of its real position,
but the contamination is confined to a small region where the
3.29~\micron\ IEF emission is faint or absent. Imaging at $L'$
(broadband 3.8~\micron) was also attempted during the same night and
at the same site, but failed due to the saturation of the array by the
high sky background at the 0\farcs295~\ppix\ scale.

Images of 1--0 S(1) \hh\ line emission at 2.12~\micron\ and of the
$\sim$2.1~\micron\ continuum emission were derived from three NIR
narrowband (1\%) filter images at 2.09~\micron, 2.12~\micron, and
2.14~\micron. These images were obtained on the Perkins 1.8~m
Telescope at Lowell Observatory, Flagstaff, Arizona. After standard
reduction procedures (sky subtraction and flat field correction), the
image of the underlying continuum emission at $\sim$2.1~\micron\ was
constructed by combining the 2.09~\micron\ and 2.14~\micron\ images,
wavelengths which are free of \hh\ emission lines \citep{MSH97,MSD99}.
To obtain an image of the 1--0 S(1) \hh\ emission line, the
$\sim$2.1~\micron\ continuum image was subtracted from the
2.12~\micron\ image after the continuum had been properly scaled to
have on average the same counts for ten unsaturated stars in the
field, which are believed not to have \hh\ emission.

In addition, \citet{LFG96} kindly provided us with their
high-resolution reduced narrowband (1\%) filter images of \ngc\ at
2.12~\micron\ and 2.18~\micron. Their images cover all of the field of
our 3.29~\micron\ image of \ngc. The images exhibit some instrumental
artifacts, notably ghost images of \STAR\ including one near
$\sim$66\arcsec~N of the star, and a row of bad pixels running
$\sim$34\arcsec~N of \STAR. We subtracted the 2.18~\micron\ image from
the 2.12~\micron\ image to create an image of the 1--0 S(1) \hh\
emission line.

We obtained images of \ngc\ at $K'$ on the University of Hawaii (UH)
2.2~m telescope at Mauna Kea, Hawaii \citep{SWD92}. Dome flats were
taken and used for the flat field correction. A sky image was
constructed from a median-filtered combination of 8 images of offsets
at 200\arcsec\ and 224\arcsec\ from \STAR. The nebula image was
constructed from a mosaic of 17 images at offsets of 0\arcsec,
30\arcsec, 71\arcsec, 100\arcsec, and 141\arcsec\ from \STAR. The sky
image was subtracted from each of the 17 nebula images before
constructing the final mosaic image. Bad pixel masks were used to
correct cosmic rays and failed pixels during the combination.

We also obtained archival {\sl Hubble Space Telescope} ({\sl HST})
images with the Wide Field Planetary Camera 2 \citep[WFPC2;][]{WFPC2}
of \ngc\ \citep*{HST,GWR00} in F606W (a wide $V$-band). Between two
available WFPC2 fields, the one centered near 56\arcsec~W 7\arcsec~N
of \STAR\ covers the field containing most nebulosity around the
northwest filaments on two of its wide-field camera chips (WF3 and
WF4). The adjacent high-resolution (PC) chip has significant
instrumental artifacts from \STAR, but these did not cross over to the
WF chips.

For comparison among different images, each was geometrically
transformed using IRAF routines to have the same orientation and scale
as those of the final 3.29~\micron\ image. After these geometric
registrations, the images were measured to be aligned with each other
better than 0\fdg1 in orientation, and scaled to the same pixel scale
to within 0.2\% in magnification (scale) difference.

\section{Images of \ngc\label{im}}

\begin{figure*}
\plotone{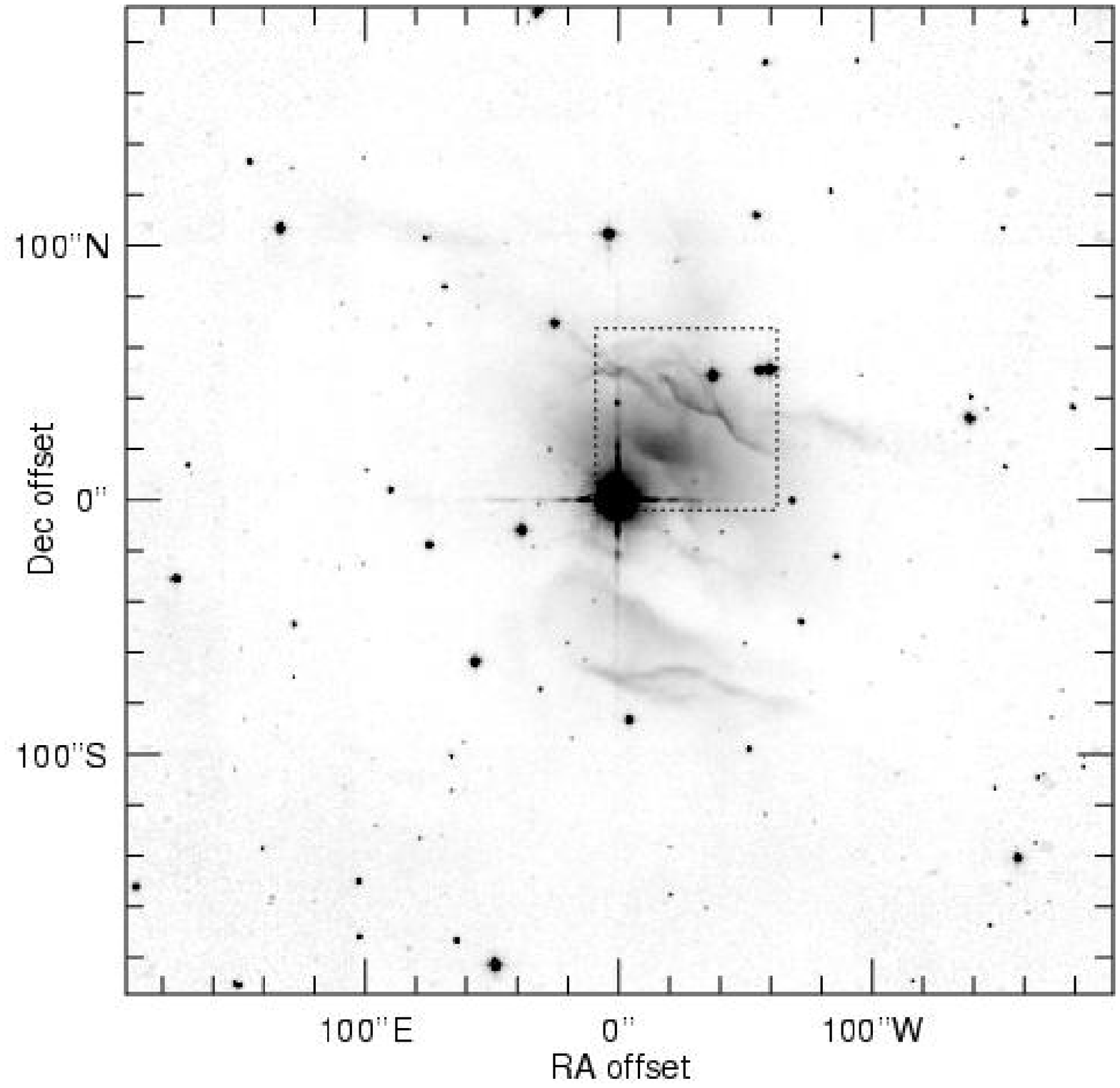}
\caption
%\figcaption[f1.eps]
{\label{f1} Image of \ngc\ at $K'$ (broadband 2.1~\micron) for a
6\farcm5$\times$6\farcm5 field centered on the illuminating star,
\STAR. The brightest nebulosity lies to the northwest of the star,
including both spatially extended emission and narrow filaments. The
rectangular box ({\it dashed lines}), centered 27\arcsec~W 32\arcsec~N
of \STAR, outlines the field covered by our 3.29~\micron\ image. The
box dimension is 71\arcsec$\times$71\arcsec, or 150~mpc $\times$
150~mpc, where 1~mpc = 1~milliparsec = 206~AU.  Seeing is 1\farcs3
(FWHM).}
\end{figure*}

Figure~\ref{f1} displays our $K'$ image of \ngc. This mosaic of images
covers 6\farcm5$\times$6\farcm5 (0.81~pc $\times$ 0.81~pc, or 810~mpc
$\times$ 810~mpc, where 1~mpc = 1~milliparsec = 206~AU), centered on
\STAR. Note that this covers more area than the $K'$ image of
\citet{SWD92}, who presented a single 3\farcm2$\times$3\farcm2
(400~mpc $\times$ 400~mpc) array image centered on \STAR\ (their
Fig.~2). The $K'$ nebulosity is mainly concentrated northwest of
\STAR, except for diffuse filamentary structures $\sim$1\arcmin\
($\sim$130~mpc) south/southwest of the star. Here, we focus our study
on the northwestern part of \ngc, represented by a rectangular box
overlaid on Figure~\ref{f1}, which corresponds to the field covered by
our 3.29~\micron\ image.

\begin{figure*}
\plotone{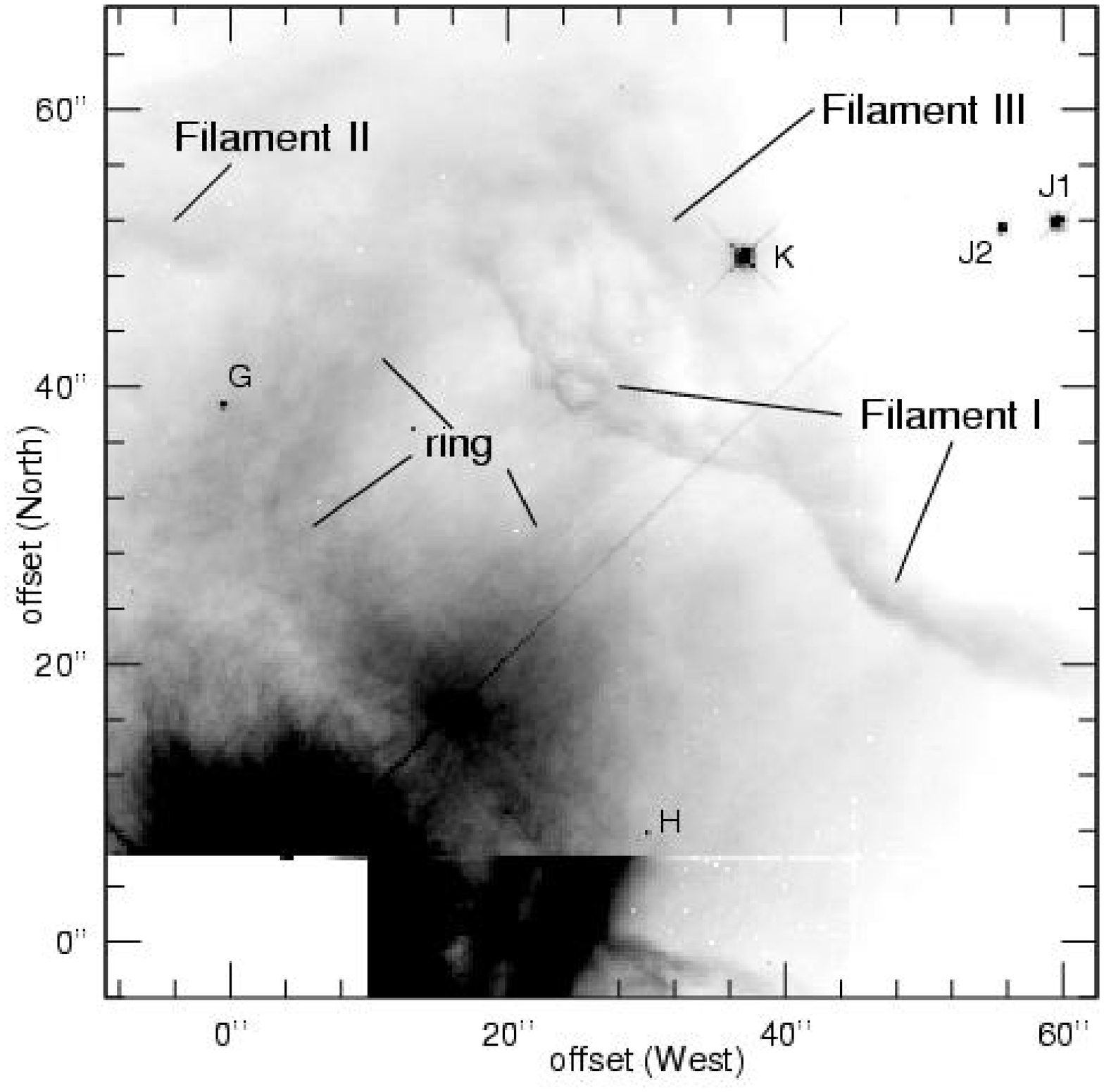}
\caption
%\figcaption[f2.eps]
{\label{f2} Archival {\sl HST}/WFPC2 F606W image \citep{HST,GWR00} of
the boxed region of \ngc\ shown in Fig.~\ref{f1}. The offsets are in
arcseconds from the illuminating star, \STAR, which lies in the region
not covered by WFPC2 chips.  Three filaments and a ring structure
discussed in the text (\S~\ref{im}), are labeled. Also indicated are
five stars identified in the field (\S~\ref{im}). On the side of the
PC chip adjacent to \STAR, instrumental artifacts are significantly
observed but they are not crossed over to the WF chips. The PSF is
$\sim$0\farcs2 (FWHM) for the PC chip.}
\end{figure*}

Figure~\ref{f2} shows the archival {\sl HST}/WFPC2 F606W image
\citep{HST,GWR00} of the northwestern part of \ngc, for the region
indicated by the box in Figure~\ref{f1}. Figure~\ref{f2} shows three
distinct filaments (labeled Filaments I, II, and III) and their
substructures, a concentrated nebulosity closer to the illuminating
star, and a diffuse ``ring''-shaped structure between them (labeled
`the ring').

In Figure~\ref{f3}, we present images of \ngc\ for the same region as
Figure~\ref{f2} obtained with different narrowband filters at
2.09~\micron, 2.12~\micron, 2.14~\micron, 2.18~\micron, and
3.29~\micron. Comparison of these images as well as $K'$
(Fig.~\ref{f1}) and optical (Fig.~\ref{f2}) images indicates that the
relative contributions of different nebular components, such as
scattered starlight, NIR continuum emission, and \hh\ line emission,
vary with wavelength and therefore do not have identical spatial
distributions.

The stars seen in Figures~\ref{f1}-\ref{f3} are predominantly pre-main
sequence stars \citep{Se83}, as is \STAR\ \citep{He60}. We detect five
point sources in the field of Figure~\ref{f2} other than \STAR. The
brightest among them is the star near 37\arcsec~W 49\arcsec~N of
\STAR\ \citep[Star K in Table IIa of][]{Se83}. Star J, as identified
by \citet{Se83}, is resolved into a visual pair separated by a
projected distance of $\sim$4\arcsec\ ($\sim$8~mpc or $\sim$1700~AU),
one near 55\arcsec~W 51\arcsec~N (J2) and the other near 59\arcsec~W
52\arcsec~N (J1) of \STAR. The stars near 1\arcsec~E 39\arcsec~N and
near 30\arcsec~W 8\arcsec~N of \STAR\ \citep[stars G and H,
respectively, after the nomenclature of][]{Se83}, were not detected by
\citet{Se83}, but are clearly observed at both optical and NIR
wavelengths (Figs.~\ref{f1}-\ref{f3}).

\subsection{the 2.12~{\boldmath\micron} {\rm 1--0 S(1) \hh} line}

\begin{figure*}
\epsscale{0.75}
\plotone{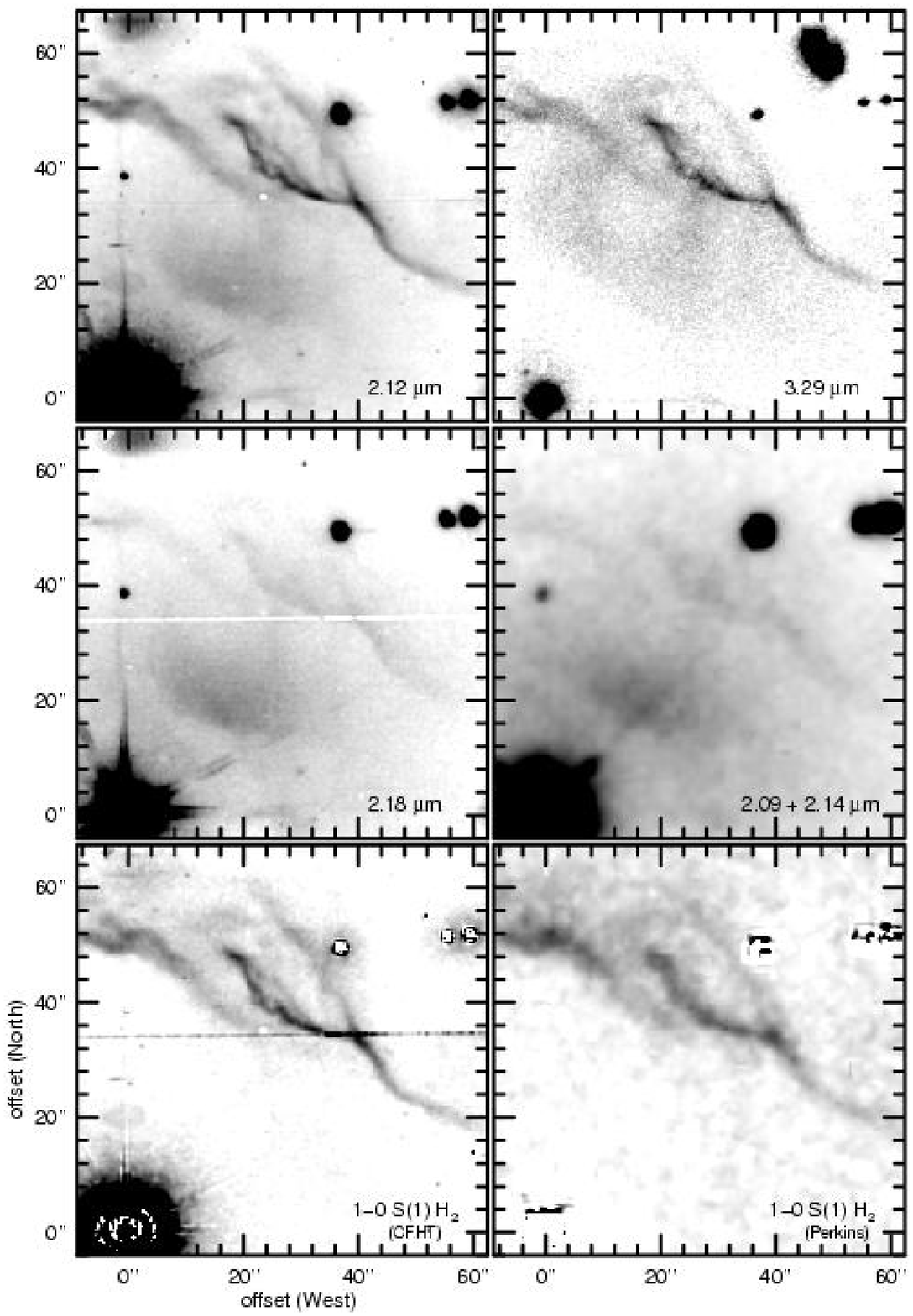}
\caption
%\figcaption[f3.eps]
{\label{f3} Narrow-band ($\Delta\lambda/\lambda$ = 1-2\%) images at
2.12~\micron, a combination of 2.09~\micron\ and 2.14~\micron,
2.18~\micron, and 3.29~\micron\ of the northwest region of \ngc. The
region shown is the same as Fig.~\ref{f2}. {\it Left column}:
2.12~\micron\ ({\it top}); 2.18~\micron\ ({\it middle}); and 1--0 S(1)
\hh\ line emission ({\it bottom}), from the difference between
2.12~\micron\ and 2.18~\micron\ images \citep{LFG96}. Seeing is
0\farcs8. {\it Right column}: 3.29~\micron\ ({\it top}),
$\sim$2.1~\micron\ continuum ({\it middle}), from the combination of
2.09~\micron\ and 2.14~\micron\ images; and 1--0 S(1) \hh\ line
emission ({\it bottom}), from the difference between our 2.12~\micron\
image (not shown) and our $\sim$2.1~\micron\ continuum image (this
paper). Seeing in our 3.29~\micron\ image is 0\farcs9 (FWHM), and in
our 2.09~\micron, 2.12~\micron, and 2.14~\micron\ images is 2\farcs8
(FWHM). The spatial structures at the upper left corner of the
2.12~\micron\ and 2.18~\micron\ images \citep{LFG96} and at the upper
right corner of the 3.29~\micron\ image are ghost images (the
instrumental reflected light of \STAR). The horizontal line,
$\sim$34\arcsec~N of \STAR, in the data from \citet{LFG96}, is due to
a row of bad pixels in their array detector. Both images of the 1--0
S(1) \hh\ emission line show artifacts due to imperfect subtraction of
the continuum for field stars.}
\end{figure*}

\citet{LFG96} have already published an image of \ngc\ in the
2.12~\micron\ 1--0 S(1) \hh\ emission line, along with images at
2.12~\micron\ (\hh\ line + continuum) and 2.18~\micron\ (continuum),
observed with CFHT. \citet{TUS00} have also published images of \ngc\
in other \hh\ emission lines. In Figure~\ref{f3}, we reproduce the
\citet{LFG96} images, as well as our new 1--0 S(1) \hh\ line image of
\ngc, for the same region as our 3.29~\micron\ image.

In both \hh\ emission line images (Fig.~\ref{f3}), the only noticeable
spatial structures are filaments northwest of \STAR, also seen in $K'$
\citep[Fig.~\ref{f1};][]{SWD92}. No significant \hh\ emission is
detected in the region between the star and the filaments, where
diffuse nebulosity appears in 2~\micron\ continuum images
\citep[Fig.~\ref{f3};][]{Se86,MSD99}. These \hh\ filaments are found
to extend over much greater lengths when observed with a wider field
of view than Figure~\ref{f3}
\citep[Fig.~\ref{f1};][]{SWD92,LFG96,TUS00}. Wider field images also
show another \hh\ filament $\sim$60\arcsec\ south/southwest of \STAR,
almost parallel to the northwest filaments shown in
Figures~\ref{f1}-\ref{f3}.

\citet{FMN96} compare their interferometric image of \HCO, which
traces dense molecular gas, with the $K'$ image of \citet{SWD92}. They
find that the \HCO\ is confined to filaments northwest of \STAR.
\citet{FMN96} show the \HCO\ filaments spatially coincide with $K'$
filaments \citep{SWD92}, while \citet{LFG96} show the $K'$ filaments
are due to fluorescent \hh\ emission, which contributes \hh\ line
emission to the broadband $K'$ filter. \citet{FMN96} detect four
kinematically separate filaments, with radial velocities between 1.9
and $4.0\ \mbox{km s$^{-1}$}$. The relationship between the
kinematically distinct \HCO\ filaments of \citet{FMN96} and Filaments
I, II, and III, as defined in Figure~\ref{f2}, is complex. Emission
from kinematically distinct filaments coincides spatially, presumably
due to projection effects, and \citet{FMN96} do not detect Filament
III in \HCO. \citet{FMN96} argue that the \HCO\ filaments are due to
increased gas density, not increased gas column density. They also
conclude that some filaments are embedded in molecular gas while
others are embedded in neutral gas.

\subsection{the 3.29~{\boldmath\micron} infrared emission feature}

Figure~\ref{f3} shows our 3.29~\micron\ image of \ngc\ along with
various narrowband images of \ngc\ near 2~\micron. The most prominent
spatial structures in the 3.29~\micron\ image are filaments which
spatially coincide with filaments in the 1--0 S(1) \hh\ images. The
3.29~\micron\ IEF is close in wavelength to the 1--0 O(5) \hh\
emission line at 3.234~\micron, but this \hh\ line falls outside the
1.5-2.0\% CVF bandpass at 3.29~\micron\ and therefore does not
contaminate our 3.29~\micron\ image.

One can easily trace three filaments in our 3.29~\micron\ image
(Fig.~\ref{f3}). Filaments I, II, and III (defined in Fig.~\ref{f2})
appear to be spatially distinct, although they do not correspond to
the four kinematically distinct filaments of \HCO\ \citep{FMN96}. The
maximum 3.29~\micron\ surface brightness (Fig.~\ref{f3}) is observed
in Filament~I. Filament~II, to the east of Filament~I and
$\sim$50\arcsec~N of \STAR, is much fainter and apparently wider at
3.29~\micron\ than Filament~I, in contrast to the \hh\ images.
Filament~III, which appears as a northern branch of Filament~I, is
much fainter at 3.29~\micron\ than the other two filaments, but it is
nevertheless clearly traceable. Much of the detailed filamentary
structure in the 3.29~\micron\ image is very similar to that seen in
images of F606W (Fig.~\ref{f2}), the 1--0 S(1) \hh\ emission line at
2.12~\micron\ (Fig.~\ref{f3}), and $K'$ (Fig.~\ref{f1}).

In addition to the filaments, diffuse nebulosity at 3.29~\micron\ is
observed closer to \STAR, forming a clear ``ring''-shaped spatial
structure at projected distances from the star of 25-50\arcsec\
(52-100~mpc). A similar structure is also seen in the optical image,
but it is much fainter and its shape is not identical to the ring in
the 3.29~\micron\ image. The diffuse nebulosity at 3.29~\micron\ is
located primarily between the filaments and the central star and is
only faintly detected farther from the star than the filaments.

Our efforts to correct the 3.29~\micron\ image for $\sim$3.3~\micron\
continuum emission, by obtaining an $L'$ image and combining it with a
2~\micron\ continuum image to interpolate a $\sim$3.3~\micron\
continuum image, were unsuccessful due to the high sky background at
$L'$. Previous studies, however, that have quantified the relative
strengths of the 3.29~\micron\ IEF emission to the underlying
continuum emission, establish that the $\sim$3.3~\micron\ continuum
contributes less than 20\% to narrowband 3.29~\micron\ photometry at
all observed positions in \ngc\ \citep{SWD83,SWA96}. We therefore
believe that our 3.29~\micron\ image is dominated by 3.29~\micron\ IEF
emission.

\begin{figure*}
\plotone{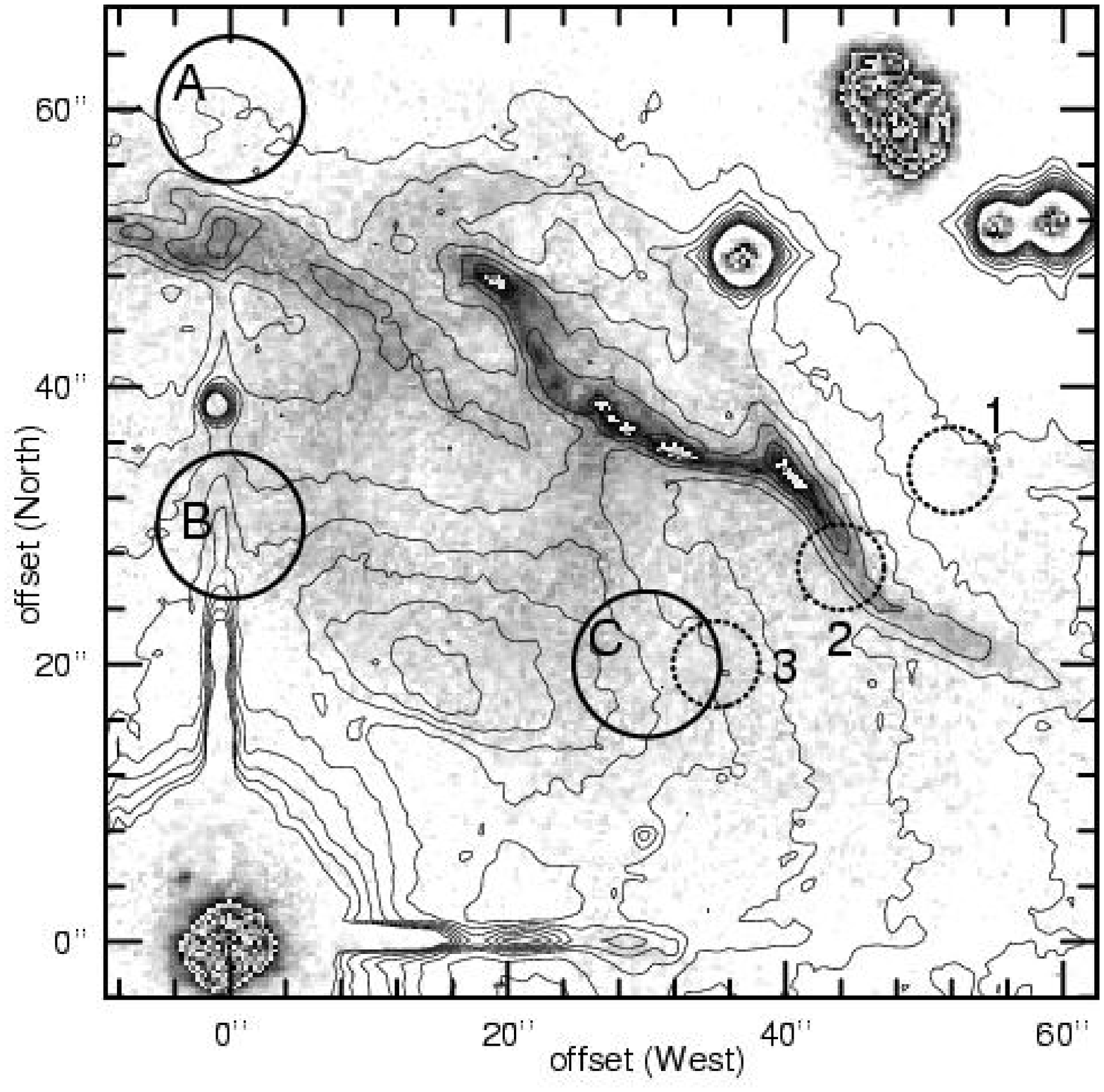}
\caption
%\figcaption[f4.eps]
{\label{f4} Contours of $K'$ surface brightness overlaid on our
3.29~\micron\ grayscale image. {\it Circles drawn with solid lines}:
the 10\farcs5 diameter apertures for both the $K$ polarimetry of
\citet{SWD92} and the $K$, 3.29~\micron, and $L'$ surface photometry
of \citet{SWD83, SWA96}. The $K$ polarization \citep{SWD92} is 13.1\%
(A), 4.6\% (B), and 4.4\% (C). {\it Circles drawn with dotted lines}:
6\farcs2 diameter aperture surface photometry at $K$, 3.29~\micron,
and $L'$ of \citet{SWA96}. The contribution of $\sim$3.3~\micron\
continuum surface brightness to the observed 3.29~\micron\ surface
brightness at these positions \citep{SWA96} is 12\% (1), 10\% (2),
17\% (3), 18\% (A), 13\% (B), and 14\% (C). The horizontal and
vertical lines in $K'$ extending from the central star to
$\sim$30\arcsec~W and to $\sim$30\arcsec~N are diffraction spikes from
\STAR, although the star $\sim$40\arcsec~N of \STAR\ is real. The
3.29~\micron\ emission structure at 48\arcsec~W 60\arcsec~N is the
instrumentally reflected image of \STAR.}
\end{figure*}

Figure~\ref{f4} illustrates positions, overlaid on our 3.29~\micron\
image, where previous photometry and spectrophotometry have quantified
the contribution of the underlying $\sim$3.3~\micron\ continuum
surface brightness to the total observed surface brightness at
3.29~\micron\ (hereafter $f$). At position C in Figure~\ref{f4}, the
3.0-3.7~\micron\ spectrophotometry of \citet{SWD83} implies $f$ =
14\%. For all other positions plotted on Figure~\ref{f4}, we derive
$f$ from surface photometry at $K$ (broadband 2.2~\micron), narrowband
3.29~\micron, and $L'$ of \citet{SWA96}. We interpolate the continuum
at 3.3~\micron\ at each position from $K$ and $L'$ data \citep{SWA96},
assuming that the continuum emission between 2.2~\micron\ and
3.8~\micron\ can be approximated by a power-law
\citep{SWD83,SAB85,SWA96}. We find that $f$ = 18\%, 13\%, 12\%, 10\%,
and 17\%, at positions A, B, 1, 2, and 3, respectively. Note that most
positions illustrated in Figure~\ref{f4} fall on diffuse 3.29~\micron\
IEF emission, with $f$ = 12-18\%, except for Position 2, which falls
on Filament~I and has $f$ = 10\%. These published observations
demonstrate that the 3.29~\micron\ IEF emission accounts for $\ga
82\%$ of the emission in our 3.29~\micron\ image.

\subsection{the 2~{\boldmath\micron} continuum}

Narrow-band images at 2.09~\micron\ and 2.14~\micron\ (shown as a
combination in Fig.~\ref{f3}), and at 2.18~\micron\ \citep[ also shown
in Fig.~\ref{f3}]{LFG96} are at wavelengths free of \hh\ emission
lines \citep{MSD99}, and thus represent the 2~\micron\ continuum
emission. Apart from differences in seeing and a few isolated
instrumental artifacts, the two images of the 2~\micron\ continuum
emission shown in Figure~\ref{f3} display a very similar morphology.
While neither 2~\micron\ continuum image in Figure~\ref{f3} has been
corrected for instrumental scattered light, they were obtained on
different telescopes, with different cameras and different
orientations of the diffraction spikes of \STAR\ on the image. Hence,
their similarity implies that the main morphological structures in
each image are intrinsic to \ngc\ and are not artifacts of the
telescope, the infrared camera, or the instrumental scattered light.

The 2~\micron\ continuum emission is a mixture of shorter wavelength
starlight transiently reprocessed by tiny particles (see
\S~\ref{nircont}) and reflected NIR starlight \citep{SWD92}.
\citet{SWD92} reported that in \ngc, the polarization rises from short
to long optical wavelengths, is strongest at $J$, and then steadily
decreases to longer NIR wavelengths, with the lowest measured
polarization values at their longest observed wavelength, $K$ (13.1\%,
4.6\%, and 4.4\% at positions A, B, and C, respectively, in
Fig.~\ref{f4}). Based on this result, they conclude that scattered
starlight contributes no more than $\sim$20\% to $K$ measurements at
positions near the 2~\micron\ continuum emission peak in \ngc. Thus,
although optical polarimetry demonstrates that the optical nebulosity
at the 2~\micron\ continuum emission peak is primarily scattered
starlight \citep{Ge67,WGS91,SWD92}, we believe that scattered
starlight makes a negligible contribution to the overall 2~\micron\
continuum emission.

The brightest nebulosity in the 2~\micron\ continuum images is the
extended peak roughly halfway between the filaments and \STAR. This
2~\micron\ continuum emission peak is centered near 17\arcsec~W
19\arcsec~N ($\sim$53~mpc) from \STAR, with a diameter of
$\sim$30\arcsec\ ($\sim$63~mpc) parallel to the filaments. The
2~\micron\ continuum emission peak roughly corresponds to the \HI\
clump detected by \citet{FMN96}. The ring structure seen in the
3.29~\micron\ image is only partially observed in the 2~\micron\
continuum images. Filament~I is clearly seen, but Filament~II is only
weakly present and Filament~III may be faintly present only in the
2.18~\micron\ image \citep{LFG96}. The filaments appear more diffuse
in the 2~\micron\ continuum emission than in the 3.29~\micron\ IEF
emission or \hh\ line emission images.

\section{Discussion\label{dis}}

\subsection{the \hh\ fluorescence compared to
the 3.29~{\boldmath\micron} infrared emission feature}

Standard PDR models \citep[see references in][]{HT97} predict that, in
PDRs, UV-pumped fluorescent \hh\ emission lines are confined to a
narrow transition region between \HI\ and \hh. Spectroscopic studies
\citep{MSH97,MSD99} and narrowband \hh\ imaging \citep{LFG96,TUS00} of
different \hh\ ro-vibrational emission lines in \ngc\ show that the
observed \hh\ emission is UV-pumped fluorescence. These authors also
find evidence that the lower level populations are redistributed
collisionally due to higher \hh\ density, especially in the \hh\
filaments, in agreement with \HCO\ results \citep{FMN96}.

Observations of the \OB, at a distance of $\sim$450~pc \citep{O2001},
show that the 3.29~\micron\ IEF emission is strongest just outside the
ionization front \citep*{ARS79,Se81,RAS89,BAW89,STN90,TMW93,SBG97}.
This has been interpreted to mean that the IEF carriers are destroyed
in \HII\ regions, perhaps by chemical attack or two-photon transient
excitation.

The peak 3.29~\micron\ emission in the \OB\ also lies significantly
closer (10-15\arcsec, or 22-33~mpc) to the star than does the peak
\hh\ fluorescent emission. A similar result is found for the planetary
nebula \PN\ \citep{GSH93}, with the 3.29~\micron\ IEF peak emission
between the \HII\ region ionization front (traced by H recombination
lines) and the \HI\,/\hh\ dissociation front (traced by fluorescent
\hh\ emission lines).

\begin{figure*}
\plotone{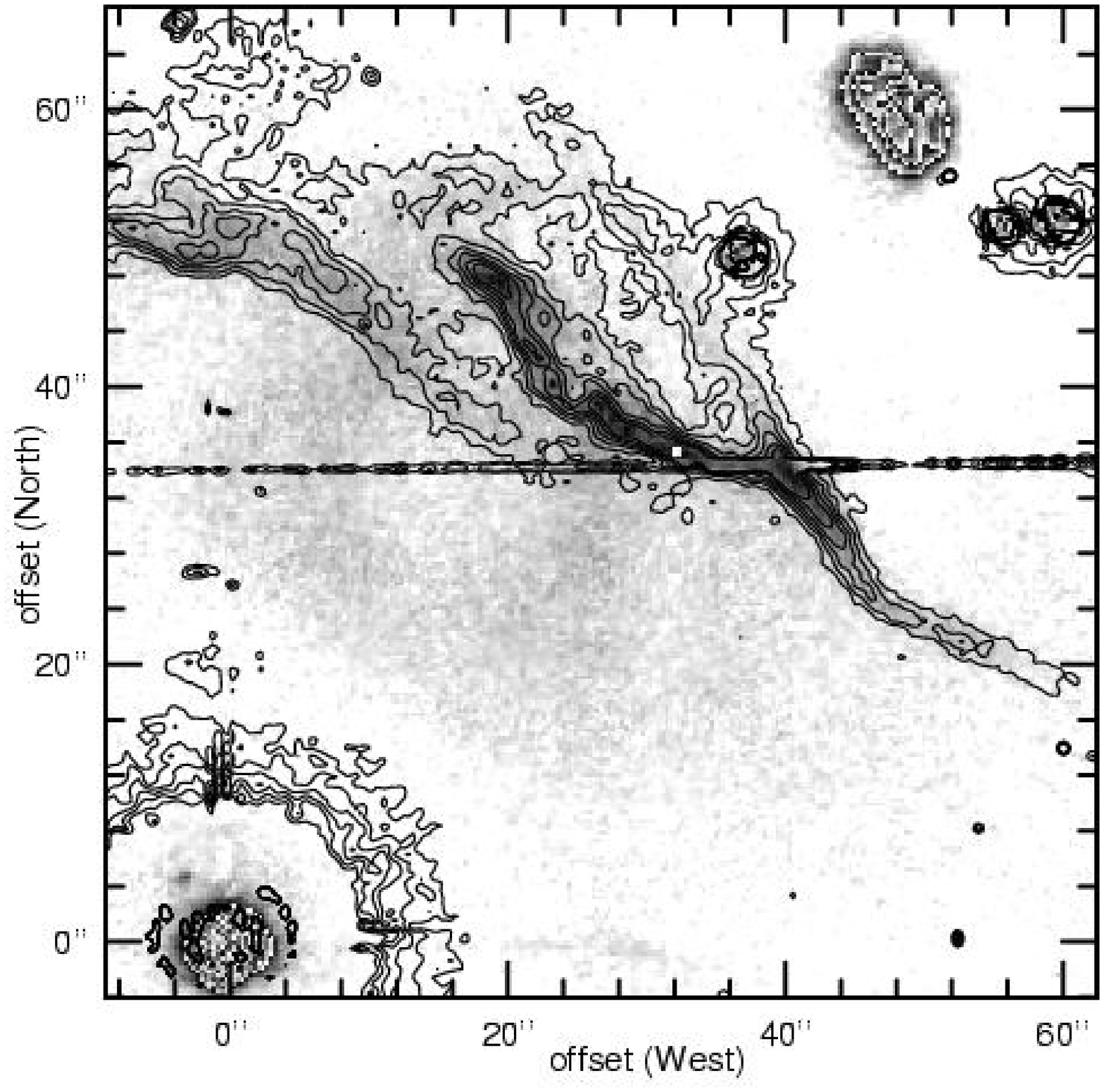}
\caption
%\figcaption[f5.eps]
{\label{f5} Contours of 1--0 S(1) \hh\ emission \citep{LFG96}, created
from the difference of 1\% spectral resolution images at 2.12~\micron\
(line + continuum) and 2.18~\micron\ (continuum), overlaid on the new
3.29~\micron\ grayscale image. Base contour (drawn 3-$\sigma$ above
the sky noise) of  the difference between 2.12~\micron\ and
2.18~\micron\ narrowband images is $I_{\nu}=6\ \mbox{MJy sr$^{-1}$}$,
and the intervals are  $3\ \mbox{MJy sr$^{-1}$}$. For 1\% spectral
resolution filters, this corresponds to a base contour of $I=8.5\times
10^{-5}\ \mbox{erg cm$^{-2}$ s$^{-1}$ sr$^{-1}$}$ and intervals of
$I=4.2\times 10^{-5}\ \mbox{erg cm$^{-2}$ s$^{-1}$ sr$^{-1}$}$ in \hh\
line emission intensity. The \hh\ filaments are coincident with the
strongest 3.29~\micron\ IEF emission, but faint 3.29~\micron\ IEF
emission is also seen between the filaments and the star.}
\end{figure*}

By contrast, Figure~\ref{f5} shows that, in \ngc, the 3.29~\micron\
IEF emission peak is essentially coincident with the \hh\ fluorescent
line emission peak, to within $\sim$1\arcsec\ ($\sim$2.1~mpc),
although diffuse 3.29~\micron\ IEF emission is also observed closer to
\STAR\ than the \hh\ fluorescent emission filaments by 15-25\arcsec\
(31-52~mpc).

The \ON\ and \ngc\ are at similar distances, and both PDRs are
believed to have more or less similar density structures, with high
density ($\geq 10^6\ \mbox{cm$^{-3}$}$) clumps embedded in lower
density ($\geq 10^4\ \mbox{cm$^{-3}$}$) interclump gas
\citep{MSH97,YO00}. The two PDRs, however, differ in three respects.
The \OB\ is illuminated by the O6 star \tCO\
(\objectname[]{HD~37022}), with \Teff\ = 40,000~K \citep{F2001}, while
\ngc\ is illuminated by the pre-main sequence B3Ve star \STAR, with
\Teff\ = 17,000~K. The higher effective temperature of \tCO\ excites
the brightest \HII\ region in Orion, the \ON, while there is no
significant \HII\ region surrounding \STAR. Finally, the UV flux
($G_0$) is ten times greater at the \OB\ than at the \hh\ filaments in
\ngc\ \citep{HT97,MSD99}.

Although detailed modeling of each PDR will be required to understand
the observed differences in the spatial distributions of the
3.29~\micron\ IEF emission relative to the \hh\ dissociation front in
the \OB\ and \ngc, it seems likely that the intensity and/or the
hardness of the incident UV field plays a major role in these
differences.

\subsection{the near-infrared continuum compared to
the 3.29~{\boldmath\micron} infrared emission feature\label{nircont}}

Previous spectra and low-resolution photometry of reflection nebulae
have shown a general association between the 3.29~\micron\ IEF
emission and the 2~\micron\ continuum emission \citep{SWD83,SWA96}
although a concrete spatial correlation (or lack thereof) between
these two types of emission has not been previously investigated.

As outlined in \S~\ref{intro}, the 2~\micron\ continuum emission is
generally thought to be the result of interactions between individual
stellar photons and large gas-phase molecules or tiny solid-state
grains, leading to transient NIR radiation. Proposed mechanisms for
the 2~\micron\ continuum emission include non-equilibrium thermal
(modified graybody) emission from stochastically heated tiny grains
\citep{SWD83,Se84}; vibrational fluorescence of many overtone bands
and combination bands of PAH molecules blended together to form a
pseudo-continuum \citep{LP84,ATB85}; continuous electronic
fluorescence from PAH molecules \citep{ATB89}; and continuous
photoluminescent electron-band transitions of solid-state particles,
most notably HAC grains or carbon nanoparticles
\citep{DW88,Du88,Du01}. Both the stochastically heated grains and the
PAH molecules are proposed to have a size of $\sim$1~nm, with 50 to
100 constituent atoms. Photoluminescence from carbon nanoparticles
would require a range of 220 to 260 atoms per nanoparticle to provide
the correct bandgap energy \citep[cf.][]{SD99,Du01} to explain the
observed 2.0-2.2~\micron\ continuum.

While the origin of the 2~\micron\ continuum emission is uncertain,
the assignment of the 3.29~\micron\ IEF to the stretching mode of
aromatic C\sbond H is generally agreed upon among the community.
Proposed laboratory analogs or identifications for the IEF carrier(s)
mainly involve hydrocarbon species: carbyne \citep{We80}; an aromatic
C\sbond H stretch \citep{DW81}; PAH molecules
\citep{LP84,ATB85,ATB89}; quenched carbonaceous composite, amorphous
carbon, or coal with varying degrees of hydrogenation
\citep*{SWT84,SWO87,BBC88,Du88,PCG89}; partially or fully hydrogenated
fullerenes \citep{We92,We93,SMC01}; nanodiamonds
\citep*{GLR99,JD00,DG01}; nonlinear \hh\ photoexcitation \citep{GS00};
and Rydberg matter \citep{Ho00,Ho01}. Neutral or ionized gas-phase
warm ($\sim$1000~K) PAH molecules \citep[see reviews by][]
{ATB89,PL89,Sa99} are currently the leading candidate carrier for the
3.29~\micron\ IEF, in the view of many researchers.

\subsubsection{the 3.29~\micron\ {\rm IEF} and 
2.18~\micron\ continuum morphology}

\begin{figure*}
\plotone{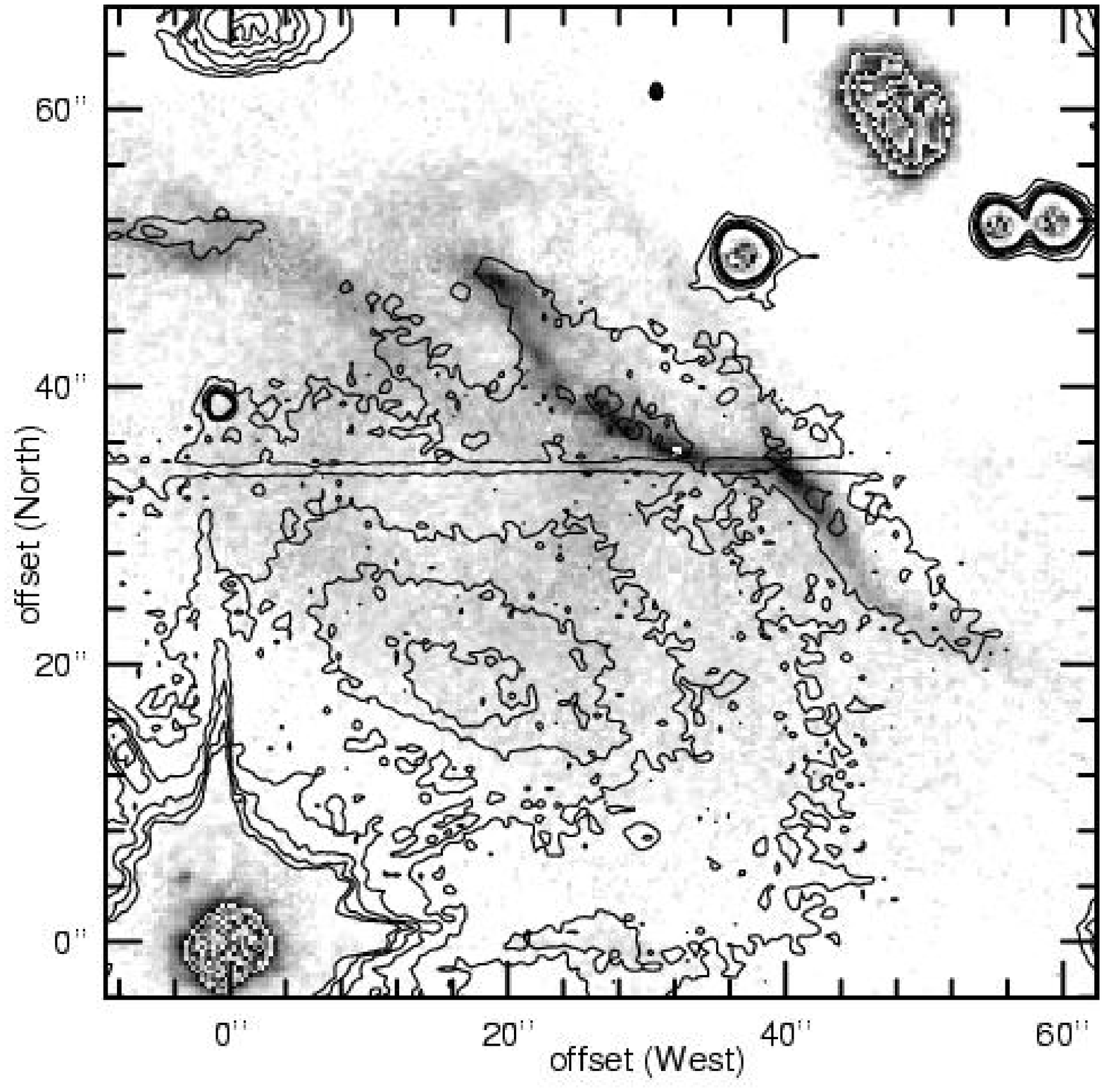}
\caption
%\figcaption[f6.eps]
{\label{f6} Contours of 2.18~\micron\ continuum emission \citep{LFG96}
overlaid on the new 3.29~\micron\ grayscale image. The 2.18~\micron\
image was obtained with 1\% spectral resolution, while the
3.29~\micron\ image was obtained with 1.5-2.0\% spectral resolution.
Thus the 2.18~\micron\ image samples the 2~\micron\ continuum between
\hh\ lines, while the 3.29~\micron\ image measures the peak emission
of the 3.29~\micron\ IEF. Base contour (drawn 4-$\sigma$ above the sky
noise) is drawn at $I_{2.18}= 6\ \mbox{MJy sr$^{-1}$}$, and the
intervals are $3\ \mbox{MJy sr$^{-1}$}$. The 2.18~\micron\ continuum
emission is brightest between the 3.29~\micron\ IEF filaments and the
star, and is weak in the 3.29~\micron\ IEF filaments. Faint
3.29~\micron\ IEF emission, however, is also seen between the
3.29~\micron\ IEF filaments and the star.}
\end{figure*}

Our high-resolution images demonstrate that the distributions of the
3.29~\micron\ IEF emission and the 2~\micron\ continuum emission are
spatially distinct in \ngc\ (Fig.~\ref{f6}). The 2~\micron\ continuum
emission is strongest in the \HI\ gas close to the star, and is
brightest at a projected distance of $\sim$50~mpc from the star,
although faint 2~\micron\ continuum emission is also seen in the \hh\
filaments at a projected distance of $\sim$100~mpc from the star. On
the other hand, the 3.29~\micron\ IEF emission is strongest in the
\hh\ filaments, with a projected separation from the peak of the
2~\micron\ continuum emission of $\sim$50~mpc. Somewhat fainter
3.29~\micron\ IEF emission, however, is also detected in the vicinity
of the peak 2~\micron\ continuum emission. This suggests that the
3.29~\micron\ IEF carriers and the 2~\micron\ continuum emitters
either are distinct and/or are excited by different physical
mechanisms.

Molecular hydrogen is both excited by and photodissociated by photons
between 91.2~nm and 110~nm. We observe both 2~\micron\ continuum
emission and 3.29~\micron\ IEF emission in the \HI\ clump near the
2~\micron\ continuum emission peak, and in the \hh\ filaments,
although their relative intensities are different. This suggests that
both the 2~\micron\ continuum emitters and the 3.29~\micron\ IEF
carriers, whether or not they are distinct, should be able to survive
in \HI\ and warm \hh.

\subsubsection{spatial dependence of the 
{\rm [}2.18~\micron{\rm ]}$-${\rm [}3.29~\micron{\rm ]} color}

\begin{figure*}
\plotone{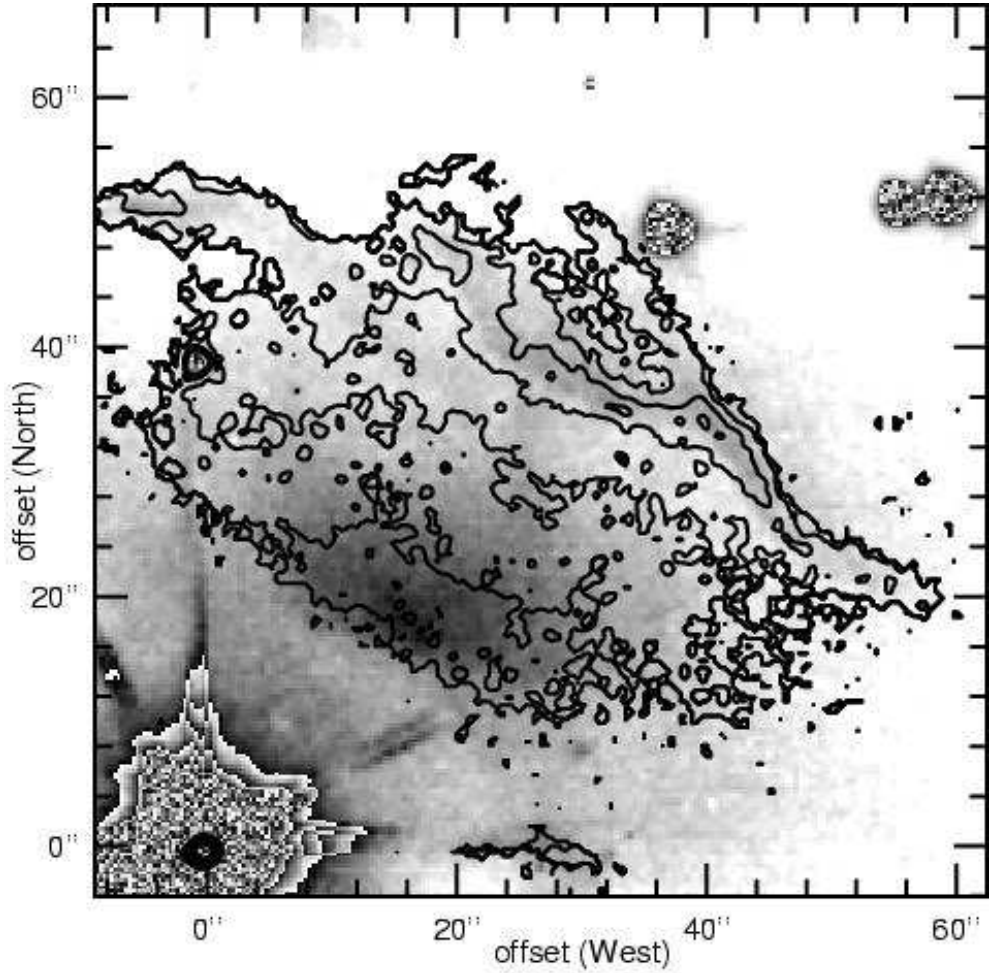}
\caption
%\figcaption[f7.eps]
{\label{f7} Contours of \ttc\ narrowband (1-2\%) colors, smoothed with
a 3~pixel $\times$ 3~pixel (0\farcs89$\times$0\farcs89) boxcar,
superposed on the grayscale 2.18~\micron\ continuum image of
\citet{LFG96}. Contour values are 3.0~mag, 3.6~mag, 4.2~mag, 4.8~mag,
and 5.4~mag, and are only shown where the emission was 3-$\sigma$
above the noise in both the 2.18~\micron\ and 3.29~\micron\ images.}
\end{figure*}

Figure~\ref{f7} quantifies the difference between the spatial
distributions of the 2~\micron\ continuum emission and 3.29~\micron\
IEF emission. We used the surface photometry of \citet{SWA96} to
derive a photometric calibration for the 3.29~\micron\ image
(Fig.~\ref{f3}), and the $K$ surface photometry of \citet{SWA96} to
derive a photometric calibration for the $K'$ image (Fig.~\ref{f1}).
We used this calibration technique in preference to the standard stars
we observed for these two images, because we were interested in the
faint extended emission on images with large sky background
subtraction and comparison with the results of \citet{SWA96}. The
2.18~\micron\ image \citep{LFG96} shown in Figure~\ref{f3} was then
calibrated by scaling it to have the same stellar magnitudes on
average for stars in common between the $K'$ and 2.18~\micron\
images. Note that the original field of view of both images are
actually larger than Figure~\ref{f3} \citep[Fig.~\ref{f1} of this
paper; Figs. 1-3 of ][]{LFG96} and the scaling is based upon six stars
common in that field. None of the stars used for scaling were
saturated.

We smoothed both the 2.18~\micron\ and 3.29~\micron\ images with a
3~pixel $\times$ 3~pixel boxcar, to increase our S/N on faint extended
emission. All images presented in this paper have been resampled to a
common scale of 0\farcs295~\ppix, which oversamples the 0\farcs9
seeing of our 3.29~\micron\ image and the 0\farcs8 seeing of the
2.18~\micron\ image \citep{LFG96}. Thus this smoothing should not
significantly degrade the spatial resolution of these two images. The
noise in the sky-subtracted, calibrated, smoothed 3.29~\micron\ image
was measured by deriving the standard deviation in regions with little
or no 3.29~\micron\ emission. All pixels with fluxes less than three
times this noise were set to zero. The same clipping process was
applied to the sky-subtracted, calibrated, smoothed 2.18~\micron\
image.

Finally, the calibrated 2.18~\micron\ and 3.29~\micron\ images were
divided, then converted to $\mbox{mag arcsec$^{-2}$}$. Figure~\ref{f7}
illustrates the distribution of the resulting color, as contours of
\ttc, overlaid on the 2.18~\micron\ grayscale image.

Figure~\ref{f7} shows the surprising result that, despite the
complicated morphology of \ngc, the \ttc\ color varies smoothly across
\ngc, with a steady increase in this color with projected distance
from \STAR, ranging from the 2~\micron\ continuum emission peak to
Filament I. Neither the 2~\micron\ continuum emission peak, nor the
``ring''-shaped region of lower 3.29~\micron\ surface brightness
between the 2~\micron\ continuum emission peak and the \hh\ filaments
(Fig.~\ref{f3}), show up as spatial structures in Figure~\ref{f7}. The
\ttc\ color illustrated in Figure~\ref{f7} reaches its largest values
in Filament I (Fig.~\ref{f2}). Observations of Filament III suggest
that the \ttc\ color begins to decrease at further projected distances
from the star, beyond Filament I.

\begin{figure*}
\plotone{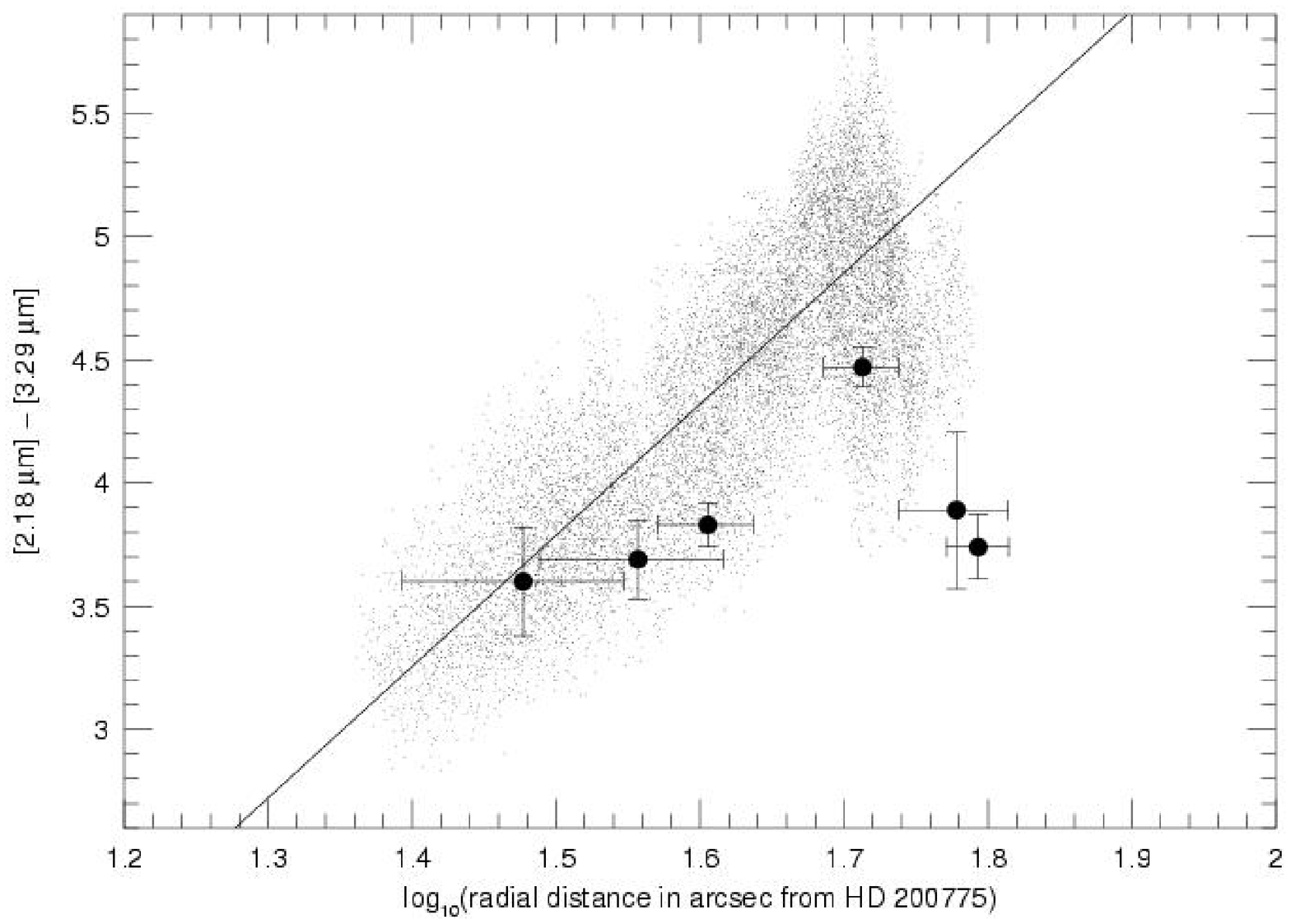}
\caption
%\figcaption[f8.eps]
{\label{f8} Individual pixel values of \ttc\ narrowband (1-2\%) colors
({\it small dots}), smoothed with a 3~pixel $\times$ 3~pixel
(0\farcs89$\times$0\farcs89) boxcar, plotted against the log of the
projected radius $r$ in arcsec from \STAR. Values are only shown where
the S/N in both [2.18~\micron] and [3.29~\micron] was $\geq 5$. Values
of $K-$[3.29~\micron] colors ({\it filled circles with error bars})
are also plotted, with the horizontal error bars representing the
photometric aperture diameter \citep{SWA96}. A least-squares fit to
\ttc\ vs.\ $\log_{10}(r)$ for $\log_{10}(r/\mbox{arcsec})<1.7$ ({\it
solid line}) is also shown, and has a slope of 5.3. This corresponds
to a surface brightness ratio of $I_{2.18}/I_{3.29}\sim r^{-2.1}$.}
\end{figure*}

The radial dependence of the \ttc\ color is quantified in
Figure~\ref{f8}. In this, we plot pixel values of \ttc\ versus
projected radius $r$ from \STAR. To reduce the scatter, we increased
the threshold for detection to 5-$\sigma$ in each \ttc\ pixel, and
masked all stars and all instrumental artifacts in the images. We also
plot, for comparison, single detector measurements of
$K-$[3.29~\micron] from \citet{SWA96}. Note that $K-$[3.29~\micron] is
generally lower than \ttc, which is in part due to contamination by
\hh\ emission lines in the $K$ filter. However, the radial pattern
followed by individual pixels of \ttc\ in Figure~\ref{f8} is generally
followed also by the lower spatial resolution data for
$K-$[3.29~\micron], including both the rise from the 2~\micron\
continuum emission peak to Filament I, and the lower values at
$r>50\arcsec$. The positions at which the $K-$[3.29~\micron] data were
taken are illustrated in Figure~\ref{f4}.

We calculated a least-square regression line to the \ttc\ pixel values
versus $\log_{10}(r)$, for $r<50\arcsec$ [or
$\log_{10}(r/\mbox{arcsec})<1.7$]. This line, with a slope of $5.33\pm
0.04$, is shown in Figure~\ref{f8}. This slope corresponds to a power
law radial dependence for the surface brightness ratio,
$I_{2.18}/I_{3.29}\propto r^{-\alpha}$, with the power index $\alpha$
being $2.13\pm 0.02$. Here $I_{2.18}$ and $I_{3.29}$ are the
2.18~\micron\ and 3.29~\micron\ surface brightnesses. This is a strong
suggestion that, in the region between the 2~\micron\ continuum
emission peak and the brightest 3.29~\micron\ IEF filament, the
difference between these two spatial distributions is related to
stellar illumination.

Neither the 2.18~\micron\ image nor the 3.29~\micron\ image is
corrected for instrumental scattered light from \STAR. Because the
spectrum of \STAR\ does not show the 3.29~\micron\ IEF \citep{Se84},
the effects of instrumental scattered light are expected to be
stronger at 2.18~\micron\ than 3.29~\micron. We investigated whether
the radial dependence of \ttc\ could be due in part to instrumental
scattered light by examining the $K'$ image (Fig.~\ref{f1}). On the
same night that the $K'$ image of \ngc\ was obtained, a $K'$ image of
\objectname[]{HR~8143} was also obtained with the same integration
time. HR~8143 \citep[$K=3.80$,][]{EFM82} is brighter than \STAR\
\citep[$K=4.62$,][]{Se83}, and is not surrounded by any nebulosity, so
should provide a robust measure of the effect of instrumental
scattered light. We corrected the $K'$ image of Figure~\ref{f1} for
instrumental scattered light, and then examined the radial dependence
of $K'-$[3.29~\micron] for $r<50\arcsec$, where \hh\ makes little or
no contribution to the $K'$ surface brightness. We found no difference
in slope for $K'-$[3.29~\micron] versus $\log_{10}(r)$ whether or not
$K'$ was corrected for instrumental scattered light. We therefore
conclude that instrumental scattered light only has minimal effects on
our observations of \ttc\ shown in Figures~\ref{f7} and \ref{f8}.

The smooth increase of \ttc\ with projected distance from \STAR\
(Figs.~\ref{f7} \& \ref{f8}) is very intriguing, given the very
different spatial distributions of 2.18~\micron\ and 3.29~\micron\ IEF
emission shown in Figure~\ref{f6}. This implies that the extended
2~\micron\ continuum emission peak is a region of higher column
density, compared to the low surface brightness region that lies
between the 2~\micron\ continuum emission peak and the \hh\ filaments.
The smooth increase in \ttc, independent of variations in column
density, therefore suggests that the main factor that contributes to
variations in \ttc\ is the decrease in stellar flux with increasing
projected distance from the star.

The radial increase in \ttc\ with increasing distance from \STAR\
could be due to several factors. We present three possibilities here,
but other explanations, such as hydrogenation or other chemical
alteration of the 3.29~\micron\ IEF carriers and/or 2.18~\micron\
continuum emitters within the \HI\ and \hh\ regions of \ngc, remain to
be explored.

It has been proposed that the 3.29~\micron\ IEF arises from the
smallest PAHs in a PAH size distribution \citep{ATB89,STA93}. It has
also been suggested that the smallest PAHs are most easily destroyed
in a strong UV radiation field \citep*{DBP90,ALS96}. If the
3.29~\micron\ IEF arises in small PAHs which are preferentially
destroyed near \STAR, and if the 2.18~\micron\ continuum arises either
from larger PAHs, from more robust tiny grains, or from larger carbon
nanoparticles, none of which are destroyed near \STAR, then this could
explain our observed radial increase in \ttc\ for \ngc.

If PAHs are the 3.29~\micron\ IEF carrier, then PAH ionization could
play a major role in the radial dependence of the \ttc\ color in
\ngc. The electron density, $n_e$, which depends on the total gas
density, is higher at the 3.29~\micron\ filaments than at the
2~\micron\ peak \citep{FMN96,LFG96,MSH97,MSD99,GPK98,TUS00}. More
importantly, the UV field, $G_0$, is also lower at the 3.29~\micron\
IEF filaments than at the 2~\micron\ peak, with $G_0 \sim r^{-2}$.
Since the PAH ionization fraction is lower for lower $G_0/n_e$
\citep{BT94}, this could result in a lower PAH ionization fraction in
the 3.29~\micron\ IEF filaments than at the 2~\micron\ peak.
Theoretical calculations and laboratory experiments show that the
3.29~\micron\ IEF is significantly weaker in ionized PAHs than neutral
PAHs \citep{dFMT93,SV93}. If the 2~\micron\ continuum emitters do not
change ionization state or their emission is not affected by
ionization state, then PAH ionization effects might be sufficient to
explain the radial dependence of \ttc\ shown in Figures~\ref{f7} and
\ref{f8}.

A third alternative has to do with extinction and the excitation of
the 3.29~\micron\ IEF and the 2~\micron\ continuum emission. The
stellar radiation from \STAR\ is systematically reddened with distance
from the star, resulting in a softer UV field at larger values of $r$.
\citet{Se84} proposed that the 2~\micron\ continuum is due to
stochastically heated tiny grains, which require energetic UV photons
to excite them. The 3.3~\micron\ IEF has been proposed to be due to
quenched carbonaceous composite, amorphous carbon, or coal
\citep{SWT84,SWO87,BBC88,Du88,PCG89}, all substances which absorb well
at visible as well as UV wavelengths. If the 3.3~\micron\ IEF could
be excited by less energetic photons than the photons required for the
2~\micron\ continuum emission, then the internal nebular extinction
within \ngc\ could cause the \ttc\ color to increase with distance
from the star. This possibility seems least likely, because
\citet{SWA96} find no evidence that the $K-$[3.3~\micron] color depends
on \Teff\ of the illuminating star, over \Teff\ = 11,000-22,000~K.

\subsubsection{observed values of 
{\rm [}2.18~\micron{\rm ]}$-${\rm [}3.29~\micron{\rm ]} in \ngc}

Another exciting result illustrated in Figures~\ref{f7} and \ref{f8}
is that the peak value of \ttc\ is $\sim$5.5~mag in the \hh\
filaments. \citet{SWA96} measured or placed upper limits on
$K-$[3.29~\micron] in 14 different reflection nebulae. In several
reflection nebulae, including \ngc, $K-$[3.29~\micron] was measured at
multiple spatial locations. The highest value of $K-$[3.29~\micron]
measured by \citet{SWA96} in any reflection nebula, over a wide range
of stellar effective temperature and stellar flux, is
$K-$[3.29~\micron]= 4.5~mag. Thus the 3.29~\micron\ IEF filaments in
\ngc\ are roughly 2.5 times brighter, compared to the 2~\micron\
continuum emission, than any other measured value in a reflection
nebula.

Care must be taken in comparing our measured values of \ttc\
(Figs.~\ref{f7} \& \ref{f8}) and published values of
$K-$[3.29~\micron]. The broadband $K$ measurements of \citet{SWA96}
include both 2~\micron\ continuum emission and \hh\ line emission,
while the 2.18~\micron\ images \citep{LFG96} were not corrected for
instrumental scattered light. We believe this has a minimal effect on
our results, since the \hh\ emission is so tightly confined to the
filaments. Values of \ttc\ greater than 4.8~mag are seen over an
extended region outside the \hh\ filaments. Another issue is the
$\sim$3.3~\micron\ continuum emission that underlies the 3.29~\micron\
IEF emission. Without an $L'$ image of \ngc, we do not know how much
of the increase in \ttc\ in Figures~\ref{f7} and \ref{f8} is due to a
redder underlying continuum emission. In particular, we cannot derive
a 3.29~\micron\ IEF feature-to-continuum ratio, which is the usual
metric of the strength of the 3.29~\micron\ IEF emission. 
\citet{SWA96}, however, presented measurements of the narrowband
3.29~\micron\ surface brightness, in their Table 4 and Figure 5, which
are not corrected for the underlying $\sim$3.3~\micron\ continuum
emission. These are the results with which we compare the data shown
in Figure~\ref{f8}.

The positions measured in the reflection nebula sample of
\citet{SWA96} were chosen by first searching for the brightest regions
of $K$ emission. The $K$ filter was chosen for initial mapping of each
reflection nebula because the high thermal background at 3.29~\micron\
makes $K$-band data vastly easier to obtain than 3.29~\micron\
narrowband data. Thus, the positions where \citet{SWA96} measured
$K-$[3.29~\micron] are biased toward being near the 2~\micron\
continuum emission peak. Figure~\ref{f7} shows that the 2~\micron\
continuum emission peak in \ngc\ has the smallest value of \ttc\ that
we were able to measure. It is intriguing that the highest
3.29~\micron\ IEF feature-to-continuum ratio measured by \citet{SWA96}
is $\sim$9, corresponding to $K-$[3.29~\micron] = 4.5~mag. Since we
measure \ttc\ = 5.5~mag in the \hh\ filaments of \ngc, we can
speculate that the 3.29~\micron\ IEF feature-to-continuum ratio could
be as high as $\sim$23 if the underlying continuum color is unchanged.
This can only be confirmed by future observations that provide an
estimate of the $\sim$3.3~\micron\ continuum emission underlying the
3.29~\micron\ IEF emission. We note, however, that Figure 6 of
\citet{SWA96} shows no correlation between $K-L'$ and 3.29~\micron\
IEF feature-to-continuum ratio. Furthermore, support for a high
feature-to-continuum ratio in the 3.29~\micron\ filaments in \ngc\ is
provided by the spectra of \citet{Joblin96}, who find 3.29~\micron\
feature-to-continuum ratios of up to 20, 34, and 37 in the reflection
nebula \objectname[]{NGC~2023}, the \OB, and the reflection nebula
surrounding NGC~1333/SVS-3 (\objectname[SVS76]{NGC~1333~3}),
respectively.

Figures~\ref{f6}, \ref{f7}, and \ref{f8} may hold out the promise of
making significant progress in understanding the emission mechanisms,
and/or the relationship between the 2~\micron\ continuum emitters and
the 3.29~\micron\ IEF carriers. The difference in the spatial
distributions of 2~\micron\ continuum emission and 3.29~\micron\ IEF
emission in \ngc, with the 2~\micron\ continuum emission strongest in
neutral gas and the 3.29~\micron\ IEF emission strongest in dense
molecular filaments, is one essential clue. Even more intriguing, the
smooth increase of the \ttc\ color with increasing projected distance
from the illuminating star of \ngc\ points towards the stellar flux
playing a crucial role in determining the relative strengths of the
2~\micron\ continuum emission and the 3.29~\micron\ IEF emission.

\section{Summary}

We compare the spatial distributions of the 3.29~\micron\ infrared
emission feature (IEF), the 2.12~\micron\ 1--0 S(1) \hh\ emission
line, and the 2~\micron\ continuum emission, in the bright visual
reflection nebula \ngc. The \hh\ emission, due to UV-pumped
fluorescence, arises in narrow, dense filaments at the dissociation
front at the edge of the molecular cloud. The 3.29~\micron\ IEF
emission is also very strong in the same filaments, but the 2~\micron\
continuum emission is quite weak there. On the other hand, the
2~\micron\ continuum emission is very bright in the region between the
\hh\ filaments and the illuminating star of \ngc, \STAR. The
2~\micron\ continuum emission is brightest $\sim$25\arcsec\
($\sim$52~mpc) from \STAR, roughly midway between the star and the
filaments. There is also a distinct ``ring''-shaped spatial structure
that emits weakly at 3.29~\micron\ and is located between the
2~\micron\ continuum emission peak and the filaments.

Our observation that both the 3.29~\micron\ IEF emission and the \hh\
fluorescent emission are brightest in narrow filaments, and are
spatially coincident to within $\sim$2~mpc, is in stark contrast to
observations of the \OB\ ionization front and of the planetary nebulae
\PN. Both the \OB\ and \PN\ show a clear spatial separation between
the 3.29~\micron\ IEF emission and the \hh\ emission. At the \OB, the
3.29~\micron\ IEF emission peak is 22-33~mpc closer to the exciting
star of the \ON\ than is the fluorescent \hh\ peak, a separation at
least a factor of ten larger than observed in \ngc. Because \ngc\ and
the \OB\ have similar gas density structures, the difference in the
projected distance between the 3.29~\micron\ IEF emission peak and
\hh\ peak between \ngc\ and the \OB\ is likely to be either related to
the intensity of the incident UV field, which is ten times higher in
the \OB\ than in \ngc, and/or related to the hardness of the incident
UV field, as indicated by the effective temperature of each exciting
star (17,000~K for \ngc\ and 40,000~K for the \ON).

We observe very dissimilar spatial distributions of the 3.29~\micron\
IEF emission and the 2~\micron\ continuum emission, and in particular,
a clear separation between the 3.29~\micron\ IEF emission peak and the
2~\micron\ continuum emission peak. This suggests three possible
interpretations, any or all of which could be correct: the excitation
mechanisms for the 3.29~\micron\ IEF emission and the 2~\micron\
continuum emission are different; the emission processes for the
3.29~\micron\ IEF emission and the 2~\micron\ continuum emission are
distinct; or the 3.29~\micron\ IEF carriers and the 2~\micron\
continuum emitters are not identical. This is in contrast to previous
assumptions that the 2~\micron\ continuum emission in visual
reflection nebula was associated with the 3.29~\micron\ IEF emission.

The spatial distribution of the 3.29~\micron\ IEF emission in \ngc\
indicates that the 3.29~\micron\ IEF carriers in \ngc\ prefer warm
molecular gas, in contrast to the separation of \hh\ fluorescent
emission and 3.29~\micron\ IEF emission in the \OB\ or \PN. The
3.29~\micron\ IEF carriers in \ngc, however, can survive in both
neutral atomic and warm molecular gas. The 2~\micron\ continuum image
indicates that the 2~\micron\ continuum emitters in \ngc\ are
predominantly associated with \HI\ gas but can survive in both neutral
atomic and warm molecular gas.

The value of \ttc, however, increases smoothly with increasing
projected distance from the star, with values of 3.4-3.6~mag at the
2~\micron\ continuum emission peak, and reaching values of 5.5~mag in
the \hh\ filaments where the 3.29~\micron\ IEF emission is brightest.
The smoothness of the radial variation in \ttc, compared to the very
structured morphology of the 2~\micron\ continuum and 3.29~\micron\
IEF images, suggests that the column densities of the 2~\micron\
continuum emitters and the 3.29~\micron\ IEF carriers are not the
origin of the differences between the spatial distributions of these
two ISM components. The smooth increase in \ttc\ with increasing
projected distance $r$ from the star, corresponding to a surface
brightness ratio $I_{2.18}/I_{3.29}\sim r^{-2.1}$, instead suggests
that the intensity of the incident stellar radiation field is directly
or indirectly related to the differences between the spatial
distributions of 3.29~\micron\ IEF emission and 2~\micron\ continuum
emission.

We do not have a specific model to explain our observation
that $I_{2.18}/I_{3.29}\sim r^{-2.1}$ in \ngc.
One idea involves the destruction of the 3.29~\micron\ IEF 
carriers, but not the 2~\micron\ continuum emitters, in the 
intense radiation field near \STAR. Another possibility
is a changing ionization fraction with distance from \STAR\
for PAHs, which have been proposed as a 3.29~\micron\ IEF carrier,
together with the assumption that the 2~\micron\ continuum
emission is insensitive to the ionization state of its
emitters or that its emitters do not change ionization state
within \ngc. A third concept invokes internal nebular extinction, 
if the 2~\micron\ continuum emitters are excited by more energetic 
photons than the 3.29~\micron\ IEF carriers.

The previously measured range of $K-$[3.29~\micron] in different
spatial positions in 14 reflection nebulae is 3.0-4.5~mag
\citep{SWA96}. Thus, the \ttc\ value at the 3.29~\micron\ IEF emission
peak is 1~mag higher than previously observed in any reflection
nebula. If $K-L'$ is independent of $K-$[3.29~\micron], as observed by
\citet{SWA96}, then the 3.29~\micron\ IEF feature-to-continuum ratio
could be as high as $\sim$23 in the \hh\ filaments in \ngc, but this
speculation remains to be confirmed by future observations.

\acknowledgements We thank the IRTF TAC for granting time for our
service observing proposal and David Griep and Charles Kaminski for
obtaining the 3.29~\micron\ images of \ngc. We thank John Rayner for
observing assistance at the UH 2.2 m. We are grateful to Alice~C.
Quillen for obtaining the 2.09~\micron, 2.12~\micron, and
2.14~\micron\ images for us. We are in debt to J.~L. Lemaire for
providing us with their published high-resolution images at
2.12~\micron\ and 2.18~\micron. We also thank B.~Scott Gaudi and
Paul~B. Eskridge for obtaining $BVR$ images for us, Paul Martini and
Darren~L. DePoy for providing us with their unpublished $JHK$ images
of \ngc, and A. Fuente for sending us electronic versions of their
published radio \HI\ data. We appreciate the careful reading of the
earlier draft of this paper by D. DePoy and A. Gould. JA thanks
B.~S. Gaudi, P. Eskridge and especially P. Martini for their help in
early reductions of some of these data. KS thanks Mark Pitts for
assistance with conversion between data formats. Most of data
reductions were performed by the use of Image Reduction and Analysis
Facility (IRAF) v2.10/2.11. IRAF is distributed by the National
Optical Astronomy Observatories (NOAO), which are operated by the
Association of Universities for Research in Astronomy (AURA), Inc.,
under cooperative agreement with the National Science Foundation
(NSF). Part of the artworks presented here were made using the OSU
implementation of XVista, which also incorporates the LickMongo
plotting package. Funding for OSIRIS was provided by grants from the
Ohio State University, and NSF grants AST-9016112 and AST-9218449.
This research has made use of the SIMBAD database, operated at Centre
de Donn\'ees astronomiques de Strabourg (CDS), Strasbourg, France.
Some of the data presented in this paper were obtained from the
Multimission Archive at the Space Telescope Science Institute (MAST).
STScI is operated by the AURA, Inc., under NASA contract NAS~5-26555.
JA was in part supported by the Presidential Fellowship from the
Graduate School of the Ohio State University. KS was supported in part
by an Alfred~P. Sloan Fellowship. This paper is dedicated by KS to the
memory of Barbara~H. Cooper (1953 -- 1999), friend and only female
classmate in graduate physics at Caltech, whose example as a brilliant
physicist helped KS to survive graduate school.

%\clearpage

%%%% Table 1

\begin{deluxetable}{cclccllcl}
\tablewidth{0pt}
\tablecaption{Observational Log\label{obs_log}}
\tablehead{
\colhead{$\lambda$ \tablenotemark{a}}&
\colhead{$\Delta\lambda$ \tablenotemark{b}}&
\colhead{Date}&
\colhead{Scale}&
\colhead{Seeing}&
\colhead{Telescope}&
\colhead{Instrument}&
\colhead{PA$_{\rm diff}$ \tablenotemark{c}}&
\colhead{Ref.}\\
\colhead{(\micron)}&
\colhead{(\micron)}&&
\colhead{(arcsec \ppix)}&
\colhead{(arcsec)}&&&
\colhead{(degr)}&}
\startdata
2.11 \tablenotemark{d}&0.35&1990 Sep 12&0.77&1.3&UH 2.2~m&
UH NICMOS \tablenotemark{e}&0&1\\
2.12&0.02&1993 Dec 1,2&0.5&0.8&CFHT 3.6~m&Redeye&0,28&2\\
2.18&0.02&1993 Dec 1,2&0.5&0.8&CFHT 3.6~m&Redeye&0,28&2\\
2.09&0.02&1995 Sep 3&1.53&2.8&Perkins 1.8~m&OSIRIS \tablenotemark{f}&45&\\
2.12&0.02&1995 Sep 3&1.53&2.8&Perkins 1.8~m&OSIRIS \tablenotemark{f}&45&\\
2.14&0.02&1995 Sep 3 &1.53&2.8&Perkins 1.8~m&OSIRIS \tablenotemark{f}&45&\\
0.586 \tablenotemark{g}&0.150&1995 Oct 4&0.10&0.20&HST 2.4~m&WFPC2&45&3,4\\
3.29&0.05-0.07&1997 Dec 13&0.295&0.9&IRTF 3~m
&NSFCam \tablenotemark{h} \ + CVF \tablenotemark{i}&0&
\enddata
\tablenotetext{a}{Central wavelength of filter.}
\tablenotetext{b}{Filter width in FWHM.}
\tablenotetext{c}{Orientation of the diffraction spikes in position
angle east of north.}
\tablenotetext{d}{The $K'$ filter.}
\tablenotetext{e}{A 1-2.5~\micron\ 256~pixel $\times$ 256~pixel
HgCdTe array detector \citep*{Hodapp92}.}
\tablenotetext{f}{A 1-2.5~\micron\ 256~pixel $\times$ 256~pixel
HgCdTe array detector \citep{DD93}.}
\tablenotetext{g}{The F606W filter for WFPC2.}
\tablenotetext{h}{A 1-5~\micron\ 256~pixel $\times$ 256~pixel
InSb array detector \citep{Rayner93}.}
\tablenotetext{i}{Circular Variable Filter of 1.5-2.0\% spectral
resolution.} 
\tablecomments{CFHT data courtesy of \citet{LFG96}.
HST data based on observations made with the NASA/ESA
{\sl HST}, obtained from the data archive at the Space Telescope
Science Institute (STScI). STScI is operated by AURA, Inc.\ under
NASA contract NAS~5-26555.}
\tablerefs{(1) \citealt{SWD92}; (2) \citealt{LFG96}; (3)
\citealt{HST}; (4) \citealt{GWR00}}
\end{deluxetable}
%\clearpage

%\centerline{\bf \textsc{Figure Captions}}\medskip

\clearpage

\clearpage

\clearpage

\clearpage

\clearpage

\clearpage

\clearpage

\end{document}